\documentstyle{mn}

\newif\ifAMStwofonts

\input epsf
\def\plotone#1{\centering \leavevmode
\epsfxsize=\columnwidth \epsfbox{#1}}
\def\plottwo#1#2{\centering \leavevmode
\epsfxsize=.90\columnwidth \epsfbox{#1} \hfil
\epsfxsize=.90\columnwidth \epsfbox{#2}}

\def\aj{AJ}                   
\def\araa{ARA\&A}             
\def\apj{ApJ}

\def\aaps{A\&AS}

\def\mnras{MNRAS}

\def\pasj{PASJ}

\ifoldfss
  \ifCUPmtlplainloaded \else
    \NewTextAlphabet{textbfit} {cmbxti10} {}
    \NewTextAlphabet{textbfss} {cmssbx10} {}
    \NewMathAlphabet{mathbfit} {cmbxti10} {} 
    \NewMathAlphabet{mathbfss} {cmssbx10} {} 
  \fi
  \ifAMStwofonts
    \ifCUPmtlplainloaded \else
      \NewSymbolFont{upmath} {eurm10}
      \NewSymbolFont{AMSa} {msam10}
      \NewMathSymbol{\upi}     {0}{upmath}{19}
      \NewMathSymbol{\umu}     {0}{upmath}{16}
      \NewMathSymbol{\upartial}{0}{upmath}{40}
      \NewMathSymbol{\leqslant}{3}{AMSa}{36}
      \NewMathSymbol{\geqslant}{3}{AMSa}{3E}

       \let\le=\leqslant
       
    \fi
  \fi
\fi 

\ifnfssone
  \newmathalphabet{\mathit}
  \addtoversion{normal}{\mathit}{cmr}{m}{it}
  \addtoversion{bold}{\mathit}{cmr}{bx}{it}
  \newmathalphabet{\mathbfit} 
  \addtoversion{normal}{\mathbfit}{cmr}{bx}{it}
  \addtoversion{bold}{\mathbfit}{cmr}{bx}{it}
  \newmathalphabet{\mathbfss} 
  \addtoversion{normal}{\mathbfss}{cmss}{bx}{n}
  \addtoversion{bold}{\mathbfss}{cmss}{bx}{n}
  \ifAMStwofonts
    \ifCUPmtlplainloaded \else
      \UseAMStwoboldmath
      \makeatletter
      \new@mathgroup\upmath@group
      \define@mathgroup\mv@normal\upmath@group{eur}{m}{n}
      \define@mathgroup\mv@bold\upmath@group{eur}{b}{n}
      \edef\UPM{\hexnumber\upmath@group}
      \new@mathgroup\amsa@group
      \define@mathgroup\mv@normal\amsa@group{msa}{m}{n}
      \define@mathgroup\mv@bold\amsa@group{msa}{m}{n}
      \edef\AMSa{\hexnumber\amsa@group}
      \makeatother
      \mathchardef\upi="0\UPM19
      \mathchardef\umu="0\UPM16
      \mathchardef\upartial="0\UPM40
      \mathchardef\leqslant="3\AMSa36
      \mathchardef\geqslant="3\AMSa3E

       \let\le=\leqslant

    \fi
  \fi
\fi 

\ifnfsstwo
  \DeclareMathAlphabet{\mathbfit}{OT1}{cmr}{bx}{it}
  \SetMathAlphabet\mathbfit{bold}{OT1}{cmr}{bx}{it}
  \DeclareMathAlphabet{\mathbfss}{OT1}{cmss}{bx}{n}
  \SetMathAlphabet\mathbfss{bold}{OT1}{cmss}{bx}{n}
  \ifAMStwofonts
    \ifCUPmtlplainloaded \else
      \DeclareSymbolFont{UPM}{U}{eur}{m}{n}
      \SetSymbolFont{UPM}{bold}{U}{eur}{b}{n}
      \DeclareSymbolFont{AMSa}{U}{msa}{m}{n}
      \DeclareMathSymbol{\upi}{0}{UPM}{"19}
      \DeclareMathSymbol{\umu}{0}{UPM}{"16}
      \DeclareMathSymbol{\upartial}{0}{UPM}{"40}
      \DeclareMathSymbol{\leqslant}{3}{AMSa}{"36}
      \DeclareMathSymbol{\geqslant}{3}{AMSa}{"3E}

       \let\le=\leqslant

    \fi
  \fi
\fi 

\ifCUPmtlplainloaded \else
  \ifAMStwofonts \else 
    \def\upi{\pi}
    \def\umu{\mu}
    \def\upartial{\partial}
  \fi
\fi

\title[6.4 keV line from Cooling Flows]{The 6.4 keV Fluorescent Iron
Line from Cluster Cooling Flows} 

\author[E. Churazov et al.]{ E.~Churazov,$^{1,2}$ R.~Sunyaev,$^{1,2}$
M.~Gilfanov,$^{1,2}$ W.~Forman$^{3}$ and  C.~Jones$^{3}$\\
$^1$ MPI f\"{u}r Astrophysik, Karl-Schwarzschild-Strasse 1, 85740
Garching, Germany \\
$^2$ Space Research Institute (IKI), Profsouznaya 84/32, Moscow 117810, 
Russia \\
$^3$ Smithsonian Astrophysical Observatory, 60 Garden St., Cambridge, MA
02138, USA}
 
\date{Submitted to MNRAS}
\journal{}

\pagerange{\pageref{firstpage}--\pageref{lastpage}}
\pubyear{0000}

\begin{document}

\maketitle

\label{firstpage}

\begin{abstract}
The fate of the cooling gas in the central regions of rich clusters of
galaxies is not well understood. In one plausible scenario 
clouds of atomic or molecular gas are formed. However the mass of the cold
gas, inferred from 
measurements of low energy X-ray absorption, is hardly consistent with the
absence of powerful CO or 21 cm lines emission  from the cooling flow region.
Among the factors which may affect  the detectability of the cold clouds are
their optical depth, shape and covering fraction. Thus, alternative methods to
determine the mass in cold clouds,  
which are less sensitive to these parameters are important. 

For the inner region of the cooling flow (e.g. within the radius of
$\sim$50--100 kpc) the Thomson optical depth of the hot gas in a massive
cooling flow can be as large as $\sim 0.01$. Assuming that the cooling
time in the inner region is few times shorter than the life time of the
cluster, the Thomson depth of the accumulated cold gas can be higher
accordingly (if most of the gas remains in the form of clouds). The  
illumination of the cold clouds by the X-ray emission of the hot gas should
lead to the appearance of a 6.4 keV iron fluorescent line, with an 
equivalent width proportional to $\tau_T$. The equivalent width only weakly
depends on the detailed properties of the clouds, e.g. on the column density of
individual clouds, as long as the column density is less than few
$10^{23}~cm^{-2}$.  Another effect also associated exclusively with the cold
gas  is a 
flux in the Compton shoulder of bright X-ray emission lines. It also 
scales linearly with the Thomson optical depth of the cold gas. 
With the new generation of X--ray
telescopes, combining large effective area and high spectral resolution, the 
mass of the cold gas in cooling flows (and it's distribution) can be measured. 
\end{abstract}

\section{Introduction}
The radiative cooling time of the gas in the central parts of rich
clusters of galaxies can be considerably shorter than the Hubble time
\cite{lea73,sil76,cb77,fn77} and the gas may cool down below
X--ray temperatures, forming a cooling flow (see Fabian, 1994 for  
review).   Strong peaks in the 
surface brightness distribution, observed in many 
clusters of galaxies (e.g. Branduardi--Raymont et al. 1981, Fabian et
al. 1981, Canizares, Stewart \& Fabian A.C. 1983, Fabricant \&
Gorenstein 1983, Jones \& Forman 1984, Stewart et al. 1984),
are usually considered as indicators of a 
cooling flow. Evidence for cool gas was also found in the
spectroscopic observations, which revealed the presence of emission lines,
characteristic of gas with the $T_e \le 5\times 10^6$ K gas (e.g. Canizares
et al., 1979). Detailed studies of the 
surface brightness distribution in the cooling flows lead to the
conclusion, that a fraction of the gas drops out from the flow at
different radii, such that the mass deposition rate is proportional to
the radius: $\dot{M} (<R) \propto R$
(e.g. Thomas, Fabian and Nulsen, 1987). The fate of the gas which has
dropped out from the
flow is not well understood. Only small fraction of the cooling gas can form
stars with a normal IMF. The optical and UV observations of the central
galaxy restrict the total rate of massive star formation rate typically by
$\sim$ 5--30 $M_\odot/year$ (e.g. Fabian et al. 1984, McNamara \&
O'Connel 1993, Cardiel, Gorgas, Aragon-Salamanca 1995, Smith et al. 1997).
Detection
of excess absorption in several cooling flow clusters (e.g. White et al.,
1991)   
seems to support the hypothesis that the cooling matter forms clouds of atomic
or molecular gas. To provide effective absorption, cold clouds
must have a covering fraction close to unity \cite{whi91}. Recent
analysis by Allen and Fabian (1997) of a sample of nearby cooling flow
clusters observed with ROSAT confirmed presence of excess absorption,
although with covering fraction somewhat lower than unity. The derived
masses of X--ray--absorbing material are  
$\sim 10^{12}~M_\odot$ \cite{whi91} and are hardly consistent
with the absence of powerful CO and HI emission (e.g. O'Dea et
al. 1994, Voit and Donahue 1995).
Thus, independent determinations of the cold gas mass are important in
understanding the fate of the cooling gas. Geometrically small,
optically thick clouds may easily escape detection. Therefore
observations at the wavelengths at which the clouds are transparent are the
best indicators of total mass of cold gas. From this point of view, X--rays
with energies above $\sim$6 keV could be useful, since clouds only become
opaque at these energies when hydrogen column density exceeds
few $10^{23}$--$10^{24}~cm^{-2}$. 

The Thomson depth of the cooling flow regions, calculated for the hot X--ray 
emitting gas, can be of the order of $0.01$. Assuming a cooling time in the
inner region which is few times shorter than the Hubble time, the  averaged
Thomson depth of the gas deposited during the life time of the cluster can
be even higher. We 
argue below that such an  amount of cold gas is sufficient to produce the
emission from the 6.4 keV fluorescent iron line with a flux well above the
sensitivity limit of future X-ray missions. Possible presence of the 6.4
keV line in clusters was also noted earlier by Vainshtein \& Sunyaev
(1980) and White et al. (1991). 
Another effect, associated with cold gas, which also might be observed
for such values of Thomson optical depth is the Compton shoulder of
the brightest X--ray emission lines. Note that (unlike CO and 21 cm
lines or low energy absorption) the fluxes in the 6.4 keV line and
Compton shoulder do not depend strongly on the shape or column density $N_H$ of
the individual cold clouds as long as  $N_H$ is lower than $\sim
10^{24}~cm^{-2}$.  

\section{Equivalent width of the 6.4 keV line}
	Let us consider the simplest case of a uniform sphere of
neutral (or molecular) gas
with an X--ray continuum source at the center. We assume that the optical
depth of the 
sphere is less than 1 for X--rays above $\sim$ 6 keV. Absorption of
the X--ray photons above the 
iron K--shell absorption edge ($E_K=7.1$ keV) will cause (with 
probability $Y\sim 0.3$, e.g. Bambinek et al., 1972) emission of 6.4
keV photons. Thus, the 
equivalent width of the 6.4 keV line for the whole region can be calculated
as (e.g. Vainshtein \& Sunyaev, 1980):
\begin{eqnarray}        \label{ew1}
EW=\frac{Y N_H A_{Fe} \int^{\infty}_{E_K} \sigma_{Fe}(E) I(E) dE}{I(6.4)}
\end{eqnarray}        
where $N_H$ is the hydrogen column density of the region, $A_{Fe}$ is the iron
abundance relative to hydrogen (we adopt $A_{Fe}=4.68~10^{-5}$,
Anders and Grevesse, 1989), $\sigma_{Fe}(E)=3.5\times 10^{-20} 
(E_K/E)^{2.8}~cm^2$ is the photoabsorption cross section (from the K shell) of
an iron atom, $I(E)$ is the incident X-ray spectrum (in
$phot/s/sm^2/keV$). 
One can rewrite equation (\ref{ew1}) in a more convenient form:
\begin{eqnarray}        \label{ew2}
EW=0.74 \left ( \frac{A_{Fe}}{4.68~10^{-5}} \right ) \frac{\int^{\infty}_{E_K}
(E_K/E)^{2.8} I(E) dE}{I(6.4)} \tau_T ~ keV
\end{eqnarray}
For hard power law spectra, the equivalent width of the 6.4 keV
line can be of 
the order of $\tau_T$~keV. In the cooling flow region, X--ray emission of the
gas is relatively soft and equivalent width may be lower 
(see however next section). We calculated the expected equivalent
width assuming, for $I(E)$, an emission spectrum of an optically thin plasma
at different temperatures (Table 
1)\footnote{Here and in the subsequent sections we used MEKA model
\cite{mgo85,mlo86,kaa92} as implemented in XSPEC v10 \cite{ar96} to
simulate thermal emission from an optically thin plasma.}. Table~1 shows that although ``hotter'' spectra have a clear
advantage in terms of producing a larger equivalent width of the iron line,
this temperature dependence is 
not too steep and a change of temperature from 10 to 1 keV results in a
decrease by only a factor of $\sim 4$ in the equivalent width.

\begin{table}
\caption{Equivalent width of the 6.4 keV iron fluorescent line, calculated
assuming an emission spectrum of optically thin plasma at
different temperatures, and normalized to unit Thomson optical depth and
solar abundance of iron (see equation \ref{ew2})} 
\begin{center}\begin{tabular}{cc}\hline \\
Temperature & EW (keV) \\
(keV) & $~~0.74~~\frac{\int^{\infty}_{E_K}(E_K/E)^{2.8} I(E)
dE}{I(6.4)}$\\ 
\\
\hline
1 & 0.24\\
2 & 0.54\\
3 & 0.71\\
4 & 0.83\\
10 & 1.09\\
\hline
\end{tabular}
\end{center}
\end{table}

\section{Thomson depth of the cooling flow region}
	To estimate the expected equivalent width one should
multiply the value, given in Table 1, by the Thomson depth of the region and
the iron abundance relative to the solar value. Assuming that all gas,
deposited in a cooling flow during a Hubble time $t_H\sim 2\times
10^{10}$~years, forms clouds of neutral gas, one
can write:  
\begin{eqnarray}        \label{tau1}
\tau_T=1.7\times 10^{-2} \left (\frac{R}{50 kpc} \right )^{-2} \left (
\frac{\dot{M}}{100 M_{\odot}/year} \right )
\end{eqnarray} 
The above estimate assumes that the cold gas is uniformly distributed
over the volume and the source of continuum radiation lies at the
center. Multiplying eq.(\ref{tau1}) by the factor from Table 1
(e.g. for $T_e=3$ keV see also discussion below) 
one can write for a final estimate of the equivalent width: 
\begin{eqnarray}  \label{ew3}
EW \sim 12 
\left ( \frac{R}{50 kpc} \right )^{-2} \left (\frac{\dot{M}}{100
M_{\odot}/year} \right )\left ( \frac{A_{Fe}}{4.68~10^{-5}}\right )~eV 
\end{eqnarray} 
Note that although the hot gas in clusters has (on average) an iron
abundance less than solar ($\sim$ 0.3--0.5), a central
concentrations of heavy elements was observed in a number of clusters
with cooling flows (e.g. Fukazawa et al., 1994). Note also that
the above estimate is insensitive to the shape of the clouds. As long as each
individual cloud has an optical depth (for Thomson scattering and for
photoabsorption above 6 keV) much less than 1, a clumpy distribution
(with an arbitrary covering fraction)  will
result in the same equivalent width as a uniform distribution. The radial
distribution of the scattering media is more important. 

\section{Discussion}
\subsection{6.4 keV line}
Let us consider the case of an isothermal cluster, i.e. whose 
emission spectrum at any place in the cluster is characterized by
the same value of temperature $T_e$. Assuming that the mass deposition
rate is proportional to the radius (e.g. Thomas, Fabian and Nulsen,
1987), one can see that the most favorable conditions for producing
6.4 keV fluorescent iron emission is in the very 
center of the cooling flow. Since the surface brightness of cooling flow
clusters is very strongly peaked, then the emission from this region may
dominate the spectrum 
even if the spatial resolution of the instrument does not allow one to fully
spatially resolve the cooling flow region. The equivalent width of the 6.4 keV
line reflects the total mass of iron in the region and is insensitive to the
shape or size of the clumps, as long as each individual clump is optically
thin for X--rays above 6 keV. On the other hand, the same clouds can be
optically thick to low energy X--rays and the efficiency of low
energy absorption can be low (recalculated per hydrogen atom). 
Thus, if clumps are dense and have a small
covering fraction\footnote{Discussion of the constraints on the cloud
parameters, which can be derived considering in detail conditions in the
cooling region(see e.g. Daines, Fabian and Thomas 1994) are beyond the
scope of this article.}, they may not contribute significantly to the
low energy  
absorption, but still effectively
produce 6.4 keV photons. Note also, that the resulting equivalent width is
almost insensitive to the temperature of the clumps, as long as the
temperature is low enough.  

\begin{figure}
\plotone{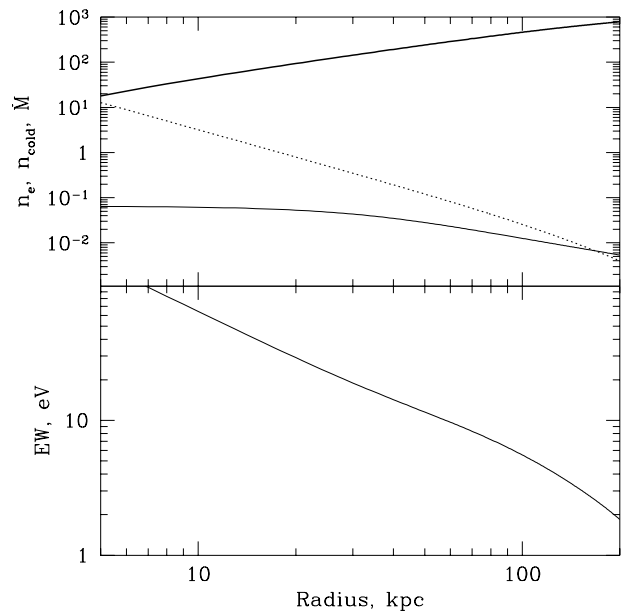}
\caption{
{\bf Upper panel:} The assumed radial distributions of hot gas density $n_e(r)$
in cm$^{-3}$ (thin solid line), mass deposition rate in $M_\odot$ per
year (thick solid line) and the density of cold gas $n_{cold}(r)$
in cm$^{-3}$ (dashed line) accumulated for $2\times 10^{10}$
years. {\bf Lower panel:} The equivalent width of the 6.4 keV line as a
function of the projected  distance $R$ from the center.   For this figure
we adopted the results on $n_e(r)$ and $\dot{M}$ of White et al. 1994.
} 
\label{a478}
\end{figure}

Allen et al. (1996) identified three distant clusters: Zwicky 3146 ($z=0.291$),
Abell 1835 ($z=0.252$) and E1455+223 as having the largest known mass
deposition rates of ~1400, 2300  and 1500 $M_{\odot}/year$. 
The equivalent width of the line from these clusters can be
of the order of  tens of eV. For nearby clusters, A478
($z=0.09$), has a large mass deposition rate of order 1000 $M_{\odot}/year$ at
$R\sim 200$ kpc (Allen et al., 1993) and the equivalent width within the 100
kpc central region may be of the order of 10
eV. One can make a ``more accurate'' estimate of the equivalent width
for this cluster 
(Fig.\ref{a478}), using the results of the deprojection analysis by Allen
et al., 1993 and White et al., 1994. Shown in the bottom panel of
Fig.\ref{a478} is the expected equivalent width of the 6.4 keV line,
calculated as a function of the projected radius from the cluster center. In
these 
calculations we assumed that the gas in the cluster is isothermal with
a temperature of $T_e\sim 3$ keV (Allen et al., 1993). Of course, given
the lack of 
knowledge of what is the real fate and distribution of the cooling gas, these
calculations may not provide more accurate estimates of the line flux
than the simple estimates in the previous section for the region as a
whole (assuming a uniform distribution of the cold gas).   

The total flux in the fluorescent line is defined by the quantity
$n_{cold}(r)\times \int^{\infty}_{E_K}(E_K/E)^{2.8} 
{\cal E}(E,r) dE$ (where $n_{cold}(r)$ is the density of cold gas and
${\cal E}(E,r)$ is the radiation density) integrated over the volume
occupied by the cold gas. Note that only that part
of the radiation density spectrum above 6 keV determines the flux
and equivalent width of the line. Although the emission measure
weighted temperature is low in the cooling flow region, compared to
the outer hot regions, it is likely that cooling flows are
inhomogeneous and hot gas is present even at small radii (e.g. Fabian
and Nulsen, 1977, Nulsen 1986, see Fabian, 1994 for review) and
above 6 keV the contribution to the radiation density from this hot phase
may be dominant (see e.g. Allen, Fabian \& Kneib 1996). One can
estimate the ``effective'' 
temperature, 
i.e. the value of $T_e$ which should be used for calculating the
line flux (see Table 1),  using the high  energy (above 6 keV) part of
the observed spectrum towards the center 
of the cluster, since the shape of the radiation density spectrum at
the cluster center and the observed spectrum along this line of sight
coincide for a symmetric cluster.  Note also, that if an AGN (having
hard X--ray spectrum) is present in the nucleus of the central galaxy,
then the equivalent width of the fluorescent line will be appropriately
higher.   

Assuming that the trend $M_{\odot}\propto R$ holds down to small radii and
cold clouds remain in the region, where they were deposited, then equivalent
width will be highest in the very central region of the cooling flow. As
noted above if the
radiation density (above 6 keV) is strongly peaked at the center of the
cooling flow then the emission 
measure weighted equivalent width from the whole region (within given
radius) can correspondingly be higher than the equivalent width calculated
for given projection radius.

\begin{figure}
\plotone{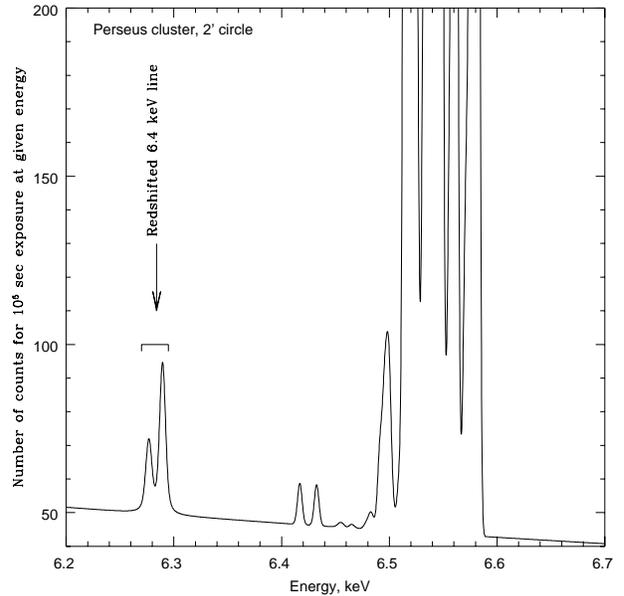}
\caption{
The number of photons, which could be detected by OXS of Spectrum--X--Gamma
at a given energy during a $10^5$ s exposure towards the Perseus cluster
(central 
2 arcmin circle). The value of 10 eV was adopted for the total equivalent width
of the 6.4 keV line (which consists of two components). The energy
resolution of the OXS is $\sim$ 5 eV. 
} 
\label{ocs}
\end{figure}

With present instruments (e.g. ASCA) the detection of the 6.4 keV line
from the cooling flow regions is questionable. The expected flux in the 6.4
keV line is 
relatively  weak compared to that of the 6.7 keV line complex and the 
combination of energy (and angular) resolution and sensitivity may not
be sufficient to detect $EW\sim$ 10 eV features from clusters. 
The better energy resolution of ASTRO-E (with bolometers),  
Spectrum--X--Gamma with the Bragg spectrometer -- OXS (see
e.g. Christensen et al. 1990)
and especially CONSTELLATION/HTXS \cite{whi97} will allow one to
detect this line from   
the clusters with massive cooling flows (if most of the cooling
gas indeed forms molecular clouds). For Perseus (assuming an
equivalent width of $\sim$ 10 eV) about 100 line photons could be
detected in a 100 ksec observation with the Bragg spectrometer of
Spectrum--X--Gamma. Shown in Fig.\ref{ocs} is the expected number of photons
which will be detected from the central $2'$ region of the Perseus
cluster. An effective area of $\sim$100 $cm^2$ (taking into account the peak
reflectivity of $\sim$21\%) and an energy resolution of $\sim$5 eV were
assumed.  
The intrinsic structure of the 6.4 keV line (which consists of two
components separated by $\sim$13 eV) was used to generate the figure. 
A few 100 ksec exposures (to measure line and continuum fluxes) should be
sufficient to detect the 6.4 keV line with an equivalent width of a few eV.
For the AXAF High Energy Transmission Grating extended nature of the source
will significantly complicate the detection of the spectral features.
CONSTELLATION with anticipated effective area of $\sim 6000~cm^{-2}$ and
an energy resolution of few eV will be able to detect very weak lines during
short exposure thus allowing studies of clusters with much weaker cooling
flows. With CONSTELLATION it should be even possible to detect 6.4 keV line
associated with column densities of the cold gas $\sim 10^{21}~cm^{-2}$
derived from low energy absorption studies \cite{whi91}.

The part of the spectrum around the 6.4 keV line should be relatively free from
contamination by bright emission lines from the hot gas. At
temperatures lower than 
$\sim$ 1 keV, excitation of the inner shells of FeXVII--FeXX ions by 
electron 
impact may produce lines at 6.42--6.47 keV, but their equivalent
widths should not be high, if the higher temperature components
(above $T_e\sim$ 1 keV) dominate the spectrum at energies of $\sim$ 6
keV.  

Along with the 6.4 keV ($K_\alpha$ ) line the $K_\beta$ line at 7.06
keV should be present with intensity about 10 times lower
($\frac{17}{150}$) than that of the $K_\alpha$ line. The absorption
edge at an energy of 7.1 keV should also be present. 
However it may be considerably more difficult to detect relatively
wide absorption feature compared to emission line. 

\subsection{Compton shoulder}
\begin{figure*}
\plottwo{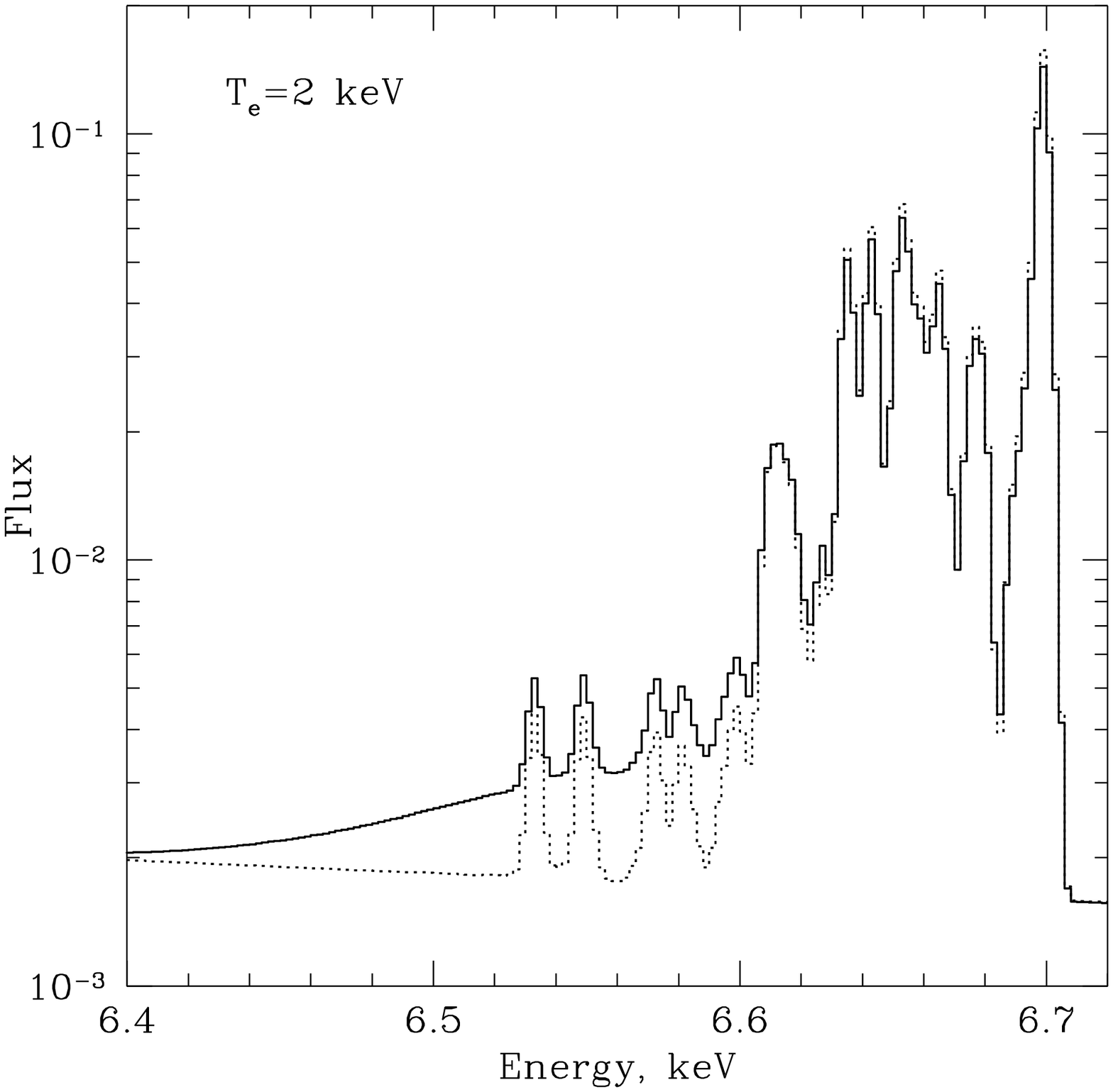}{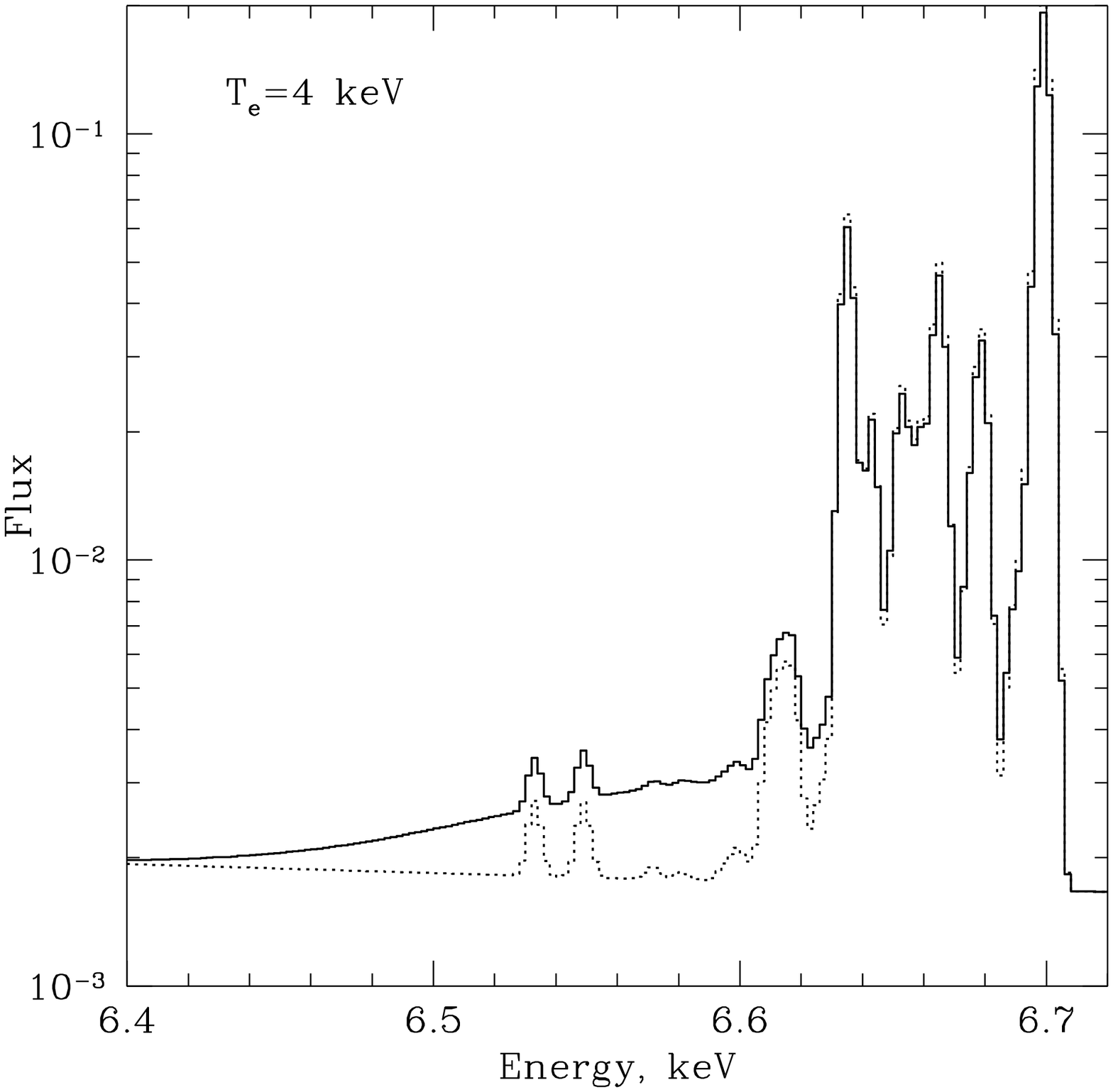}
\caption{
The dotted line shows the emission spectrum from an  optically thin plasma
(MEKA 
model as produced by XSPEC) at $T_e=2$ keV (left) and  $T_e=4$ keV
(right), convolved with a FWHM=5 eV Gaussian. Solid line shows 
the effect on the spectrum of scattering by neutral hydrogen (for a Thomson
optical depth of 0.1). The scattered $\sim$ 6.6--6.7 keV
photons have slightly lower energies, causing an increase in the
continuum level at 6.5--6.6 keV by a factor $\sim 1.5\times
(\frac{\tau_T}{0.1})$). Note that in the presence of lower
temperature components this 6.5--6.6 keV region may also be
contaminated by weaker emission lines. 
} 
\label{compt}
\end{figure*}

Another effect, which also scales linearly with the Thomson depth of
the cold gas is the Compton shoulder, associated with the brightest
emission lines. Assuming a Thomson depth for the cold gas of $\tau_T\sim
0.1$ one can conclude that $\sim$ 10\% of all photons (in particular, those
in emission lines) will be scattered, forming the so-called Compton
shoulder at the left (low energy) side of the emission lines
(Fig. \ref{compt}). Hot gas does 
not contribute to the formation of the Compton shoulder, since large thermal
velocity of the electrons causes complete smearing of the features. On
the other hand scattering of the emission lines by electrons bound in
atoms and molecules will produce the Compton shoulder
with distortions, specific to the given type of atom or molecule
\cite{sc96}. This effect is also weakly dependent on the cloud
parameters and in principle provides the unique possibility of
distinguishing between the 
contributions of free ($\sim$ cold) electrons, atomic hydrogen, atomic
helium or singly ionized helium, although presence of blends of
emission lines (like the 6.7 keV complex) may significantly complicates
the measurements. In principle, combined measurements of the iron
$K_\alpha$ line and the Compton shoulder allow one to determine the
iron abundance ($\frac{Fe}{H}$) in the cooling flow.

\section{Conclusions}
	Future X--ray missions, combining high energy resolution and
large effective area, will be able to estimate the amount of cold gas
deposited in  massive cooling flows, by measuring the equivalent width
of the 6.4 keV iron fluorescent line or the intensity of the Compton
shoulder, associated with brightest X--ray emission lines. 
Assuming that the bulk of the cooling gas forms  
cold molecular clouds, the expected fluxes are high enough to ensure
their detection. These measurements could provide robust estimate of
the gas mass, insensitive to the detailed shape, density or
temperature of the cold clouds, as long as column densities of
individual clouds are less than $\sim 10^{24}~cm^{-2}$. In
particular, the intensity of the 6.4 keV line will be less 
dependent on the covering fraction of the clouds than either the magnitude of
the low energy absorption or the 
intensity of the 21 cm or CO lines, both of which are more sensitive to
the column densities of individual clouds.

This work was supported in part by the RBRF grant 
96-02-18588. C. Jones and W. Forman acknowledge support from the
Smithsonian Astrophysical Observatory and AXAF Science Center.

\label{lastpage}

\end{document}